# Evaluating the Effect of Timeline Shape on Visualization Task Performance

**Sara Di Bartolomeo** , **Aditeya Pandey** , **Aristotelis Leventidis** ,
**David Saffo** , **Uzma Haque Syeda** , **Elin Carstensdottir** ,
**Magy Seif El-Nasr** , **Michelle A. Borkin** , **Cody Dunne**
Northeastern University
(dibartolomeo.s | pandey.ad | saffo.d | syeda.u)@husky.neu.edu, elin@ccs.neu.edu,
(magy | m.borkin | c.dunne)@northeastern.edu

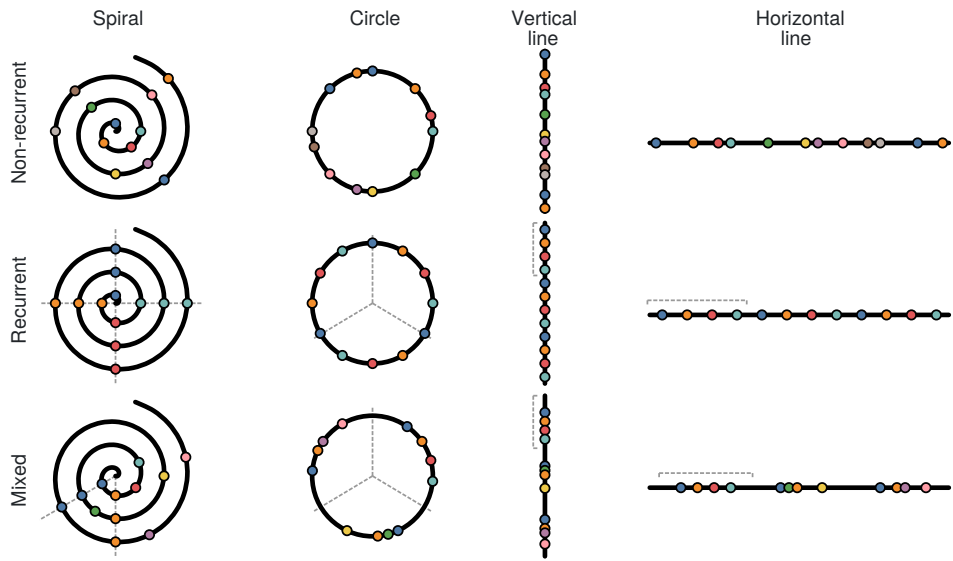

Figure 1. We evaluate the effect on task performance of 4 timeline shapes (left to right) across 3 types of temporal event sequence data (top to bottom). These images are simplified versions of the stimuli that we used in our experiment. Each dot on a timeline represents an event and has a specific categorical color to highlight where the dataset has recurrent events. Dashed lines highlight the recurrent intervals or a set of recurrent events.

## ABSTRACT
Timelines are commonly represented on a horizontal line, which is not necessarily the most effective way to visualize temporal event sequences. However, few experiments have evaluated how timeline shape influences task performance. We present the design and results of a controlled experiment run on Amazon Mechanical Turk ($n = 192$) in which we evaluate how timeline shape affects task completion time, correctness, and user preference. We tested 12 combinations of 4 shapes — horizontal line, vertical line, circle, and spiral — and 3 data types — recurrent, non-recurrent, and mixed event sequences. We found good evidence that timeline shape meaningfully affects user task completion time but not correctness and that users have a strong shape preference. Building on our results, we present design guidelines for creating effective timeline visualizations based on user task and data types. A free copy of this paper, the evaluation stimuli and data, and code are available at https://osf.io/qr5yu/

## INTRODUCTION
A timeline is a visual representation of a series of events in time. The use of timelines dates back to 17th century [32] when Joseph Priestly designed a visualization that showed the rise and fall of empires in Europe's history. In the modern era, timelines have become prevalent in our daily lives as the de facto representation to show financial trends, weather details, and meeting schedules. Timelines are most commonly drawn linearly [6], where the events are organized along a straight line. In practice, however, we can find abundant examples of timelines where events are arranged in non-linear shapes like circles, spirals, grids, and other arbitrary arrangements [6].

The visualization literature provides sufficient evidence that the layout and orientation of visualizations affect user's analytical task performance [14, 18]. In a classic study, Cleveland and McGill [14] and later Heer et al. [18] studied how people decode data presented to them using different shapes and orientation. Their results show that visual representation affects human ability to accurately read the data. Extrapolating from previous research in Information Visualization, we argue that there might be different perceptual effects of representing temporal event sequence data with varying timeline shapes. However, existing work in timeline visualization evaluation has not measured the impact of timeline shape alone on user task performance for general temporal event sequence data.

Timelines represent temporal event sequence data. In a review of related work and discussion with experts working with such data in the fields of history and personal health informatics, we identified 3 types of temporal event sequence data: (1) A non-recurring series of events, e.g., world history where events do not repeat. (2) A recurring sequence of events where the events happen at specific intervals, e.g., a company's quarterly financial reports. (3) A mixture of recurring and non-recurring sequences. Our lives are the best example of the third category, where certain events are recurring like time of meals, whereas other events are not like the birth of a child.

Intuitively, we can reason that a circular timeline may help readers notice a repeating pattern in the data, while a non-recurrent type of dataset may be best represented on a line to emphasize the linearity. However, there are no existing studies which systematically enquire whether the timeline shape effects reading time and accuracy for different types of temporal event sequence data.

Analytical tasks also play a role in a user's ability to read visualizations [10, 28, 33]. We interviewed two experts to create a list of common tasks on temporal event sequence data. We found that in a general day-to-day setting the experts used timelines for 4 analytical tasks: (1) a When task to identify the time associated with an event, (2) a What task to identify an event associated with a time, (3) a Find task to spot a data point on timeline based on both time and event, and (4) a Compare task which requires finding the distance between events.

In this paper, we present the first study which evaluates the readability of timeline shape alone on user task performance for general temporal event sequence data. In a crowd-sourced experiment, we compare 4 timeline shapes — horizontal line, vertical line, circle, and spiral — using 3 types of temporal data — recurrent, non-recurrent, and mixed. Our study is carefully designed to evaluate timeline shapes using common everyday tasks with familiar-looking datasets. E.g., find the date associated with an historical event on a timeline or lookup your daily schedule to find what are you supposed to do tonight at 8pm. In a within-subjects study design, we measured time to complete a visualization task and the task accuracy across the 4 timeline shapes.

We found evidence that task completion time is dependent on the choice of the timeline shape. Specifically, linear shapes were on an average faster to read. Our quantitative results are backed by qualitative feedback, where we see a strong preference for linear shapes among the participants.

**Contributions:** We contribute an overview of common timeline visualization tasks. We use these tasks to design a crowd-sourced experiment which compares readability across 4 common timeline shapes — horizontal line, vertical line, circle, and spiral. Based on our results, we recommend timeline design considerations for people interested in visually presenting temporal event sequence data.

**RELATED WORK**

Visualization of temporal event sequence data has a large body of literature. Existing work lays a strong emphasis on applied research of timeline visualizations and creating novel ways to visualize temporal event sequence data. In our review of related work, we summarize the research on timeline shape and how shape is used in practice. Unlike the design of timelines, evaluation of timelines is still in the early stages with only a few empirical studies evaluating the design of timeline shapes. In this section, we also discuss existing research evaluating timelines and identify areas of knowledge gap.

**Timeline Visualizations and Tasks.** Timeline visualizations come in a variety of shapes and are designed with a multitude of encodings and data. Brehmer et al. [6] present a design space for timelines used in storytelling. This design space identifies 5 representations (shapes) timelines take: linear, radial (circle), grid, spiral, and arbitrary. Aigner et al. [1] present a comprehensive book covering the visualization of time-oriented data. Much of the material covered is directly related or relevant to timeline visualization. They identify two possible arrangements of data: linear and cyclical.

In addition, Aigner et al. [1] include a review of several timeline visualizations with varying shapes, data arrangements, and encodings. These include examples of linear [2, 27], circular [13, 20], and spiral [11, 36] timeline visualizations. In each of these cases the timeline shape was chosen to best emphasize certain features of the data or to assist specific visualization tasks. E.g., linear timeline visualizations are generally used for navigation of the data or to display a linear sequence of events. However, circle and spiral timeline visualizations are used to highlight the cyclical or serial-periodic nature of data — such as the seasons of the year. Aigner et al. [1] also characterize several user tasks that are commonly performed with temporal data. These tasks included finding at what time events happened, finding what events happened at a specific time, comparing the interval of time between two events, identifying groups of related events, and more.

**Timeline Evaluation.** Previous work has evaluated the utility of various timeline visualizations for specific tasks. Schwab et al. [30] evaluated how quickly users can navigate to a particular data point on a timeline using pan and zoom techniques.

Brehmer et al. [7] compared linear vs. circular layout timelines on mobile devices on their efficacy for showing ranges. Recently, Waldner et al. [35] published a study on timeline shape focused on comparing juxtaposed radial charts vs. horizontal linear bar charts and two juxtaposed 12-hr charts vs. a single 24-hr chart. Their design intrinsically measured the combination of shape and juxtaposition technique. Our work complements Waldner et al.'s by evaluating the effect of simple timeline shapes alone and for additional shapes. Moreover, the data used by Waldner et al. was limited to a 24-hour setting while we test events more generally over different time scales. Furthermore, they used quantitative data associated with the occurrence/frequency of events while we focus on events only so as to measure effects for different types of event sequences.

Apart from empirical evaluations, researchers have evaluated the usability of various timeline shapes. Larsen et al. [22] found in a qualitative user study that spiral timelines were effective in identifying cycles in the data. Separately, Nguyen et al. [25] evaluated a novel linear timeline layout and found the layout to be easy to learn and effective for sensemaking.

## RESEARCH QUESTION AND HYPOTHESES

Our primary motivation is to study how different timeline shapes affect a user's ability to read data with a timeline visualization. To answer this research question, we measure a user's ability to perform a visualization task in terms of their accuracy and completion times. Further, we also take into account their overall preferences when they are performing visualization tasks. We present the formal questions which compare these measures, as the following hypotheses:

1. We will not observe a substantial difference in time or accuracy between timeline shapes in any dataset.
2. We will observe higher user preference and confidence with linear timelines for all datasets.

Based on our interviews with experts, literature review, and intuitions as visualization researchers, we expect the time difference between timeline shapes to be inconsequential for all tasks and questions. Timeline shape will not be detrimental to participants' ability to complete tasks in a timely manner. However, we expect users to show a preference towards linear vertical and horizontal timelines as these are the most conventional.

## STUDY

The stimuli used in this study were carefully and purposefully designed, discussed, and refined. In this section, first we motivate and explain our stimuli design. After stimuli description, we present our experimental methods. The experiment design measures the effect of the independent timeline design variable, on the dependent accuracy and time variables for 4 visualization tasks. Here, we walk you step-by-step through the experiment design and procedure.

**Stimuli**
**Datasets.** After careful inspection of the literature related to timelines and their use [1, 6], we identified 3 major types of data that are represented using timelines: non-recurrent, recurrent, and a combination of both (mixed). A non-recurrent dataset has distinct entries and does not repeat itself whereas a recurrent dataset has entries that repeat after a certain interval. A mixed dataset has both these properties, i.e, it contains both distinct and repeating entries. We created a dataset for each of these characteristics using fictitious data. This was done to avoid the potential for participants to be biased by previous knowledge of real data. For example, a participant may already know the sequence of events in a 'history of Egypt' dataset. This would give them an advantage over other participants and potentially skew results. Fictitious data avoids this problem and ensures participants will have no prior knowledge of the event sequences. Our 3 datasets — history (non-recurrent), gardening (recurrent), and schedule (mixed) — are demonstrated in Table 1.

| Datasets | | |
|---|---|---|
| **History: Non-recurrent** | **Gardening: Recurrent** | **Schedule: Mixed** |
| 912, Major epidemic | Spr '18, Dittany | Mon 8AM, Wake up |
| 915, Population plummets | Sum '18, Gurdyroot | Mon 12PM, Lunch |
| 917, Tribes migrate | Aut '18, Puffapod | Mon 4PM, Social Hour |
| 918, They settle | Win '18, Wolfsbane | Mon 8PM, Gym |
| 921, Farming starts | Spr '19, Dittany | Tue 8AM, Wake up |
| 924, Agriculture improves | Sum '19, Gurdyroot | Tue 10AM, Reading Group |
| 925, Bartering ends | Aut '19, Puffapod | Tue 12PM, Lunch |
| 927, Temples built | Win '19, Wolfsbane | Tue 6PM, Movie Night |
| ... | ... | ... |

Table 1. A overview of the datasets used in the experiment. The colors except black indicate recurrent events. (a) A non-recurrent historical timeline with invented dates and corresponding events. (b) A recurrent planting schedule of 4 different invented plants. (c) A mixed dataset consisting of an invented schedule with both repeating and non-repeating events.

**Overview of Visual Stimuli.** We use the design space proposed by Brehmer et al. [6] to finalize our experimental stimuli. Brehmer et al. divide the timeline design space into 5 categories: Linear, Radial, Grid, Spiral, and Arbitrary. We evaluate 3 of the 5 shapes: Line, Radial (called Circle in this paper), and Spiral. We exclude Grid because it is not a line-based representation, and Arbitrary because it is difficult to assign a single shape given its multiple forms. We ultimately developed 4 timeline shapes: a horizontal line −, a vertical line |, a circle ○, and a spiral ◉. Our timeline design aims to maximize the effect of timeline shape and reduce the impact of other visual embellishments. As a result, our timelines do not use any form of color-coding. Only ticks and textual labels are used to represent events. All the labels are horizontal. To support comparison across the shapes, we keep the font size and style consistent across the designs. The final timelines are black lines with ticks indicating the position of events. Each event is further represented by a label for the date and another one for the name of the event. Events on the timeline are positioned at distances proportional to their distance in time. Each timeline is drawn within a pre-determined frame size which is rendered using 70% of the user's screen size. The timelines were implemented in JavaScript using D3 [5].

**User-Centered Timeline Design.** A timeline design has concerns beyond its visual presentation. E.g., how do people read timelines? We did not want to make an assumption about the readability of the timeline based on our intuition. To understand how people read the 4 timeline shapes used in our experiment, we conducted an in-person survey with 11 partic-ipants from Northeastern University, Boston. In the survey, we asked the participants how they will read each of the 4 timelines. Our findings, summarized in Table 2, guided our stimuli design for the full study.

| Shape | Way to read | Starting point |
|---|---|---|
| — | Left to right | Left |
| \| | Top down | Top |
| ○ | Clockwise | Top |
| ◉ | From the inside | Top, going to the right |

Table 2. Results of timeline readability survey. Here we show the most common readability technique that emerged for each timeline shape.

We used a similar approach to find the maximum number of points that can be shown on each shape without making it look overwhelming or cluttered. We built a web application with sliders to adjust the font size and number of points while simultaneously showing the effect of the changes on each of the shapes. We then asked participants to adjust the sliders until they reached the maximum number of points that still felt comfortable to read. Based on our findings, we decided to display 12 data points on each one of the visualizations.

**Final Stimuli Designs.** We compare the readability of 4 timeline shapes with 3 temporal event sequence datasets. To test each shape with each dataset, we represent all 3 types of datasets on all 4 timeline designs. This led to 12 unique combinations of timeline shape and dataset. We present an overview of these shapes in Figure 1.

**General Design Patterns.** Key design elements were consistent across all 4 timeline shapes. Each timeline contains 12 data points and their dimensions are proportionally regular to account for varying screen sizes of participants. Each timeline has arrows denoting the beginning and end. Events are displayed with two labels: name of the event and date.

**Dataset-Specific Design Patterns.** There are some visual concerns related to the representation of each dataset with each shape. One of these is to aid readability between two labeled events. To support the users in tracking time difference between two labeled events, we use small ticks to display units of time between the events. The tick is placed according to the dataset: for the history dataset each tick denotes one year, for the gardening dataset each tick denotes one season, and for the schedule dataset each tick denotes two hours.

**Horizontal and Vertical Line.** Figure 2a shows the history timeline represented on a horizontal line — . We alternate the vertical positioning of the labels to avoid the text overlapping. The vertical line \| history timeline is shown in Figure 2b. In both a difference in time is represented as the same Euclidean distance along the line.

**Circle and Spiral.** On circle and spiral, a difference in time along the curve is represented by the angle difference from the center. For instance, if two events are separated by one year then the separation between them be an equal angle. In Figure 2c, we show how 3 equal-angle sectors preserve the distance between 3 recurrent events. In the spiral, preserving the

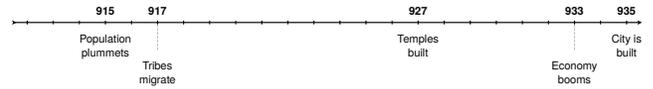

(a) A non-recurrent dataset on a horizontal line — timeline.

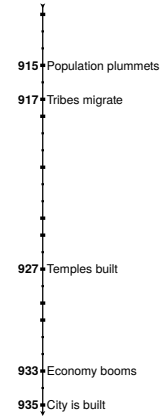

(b) A non-recurrent dataset on a vertical line \| timeline.

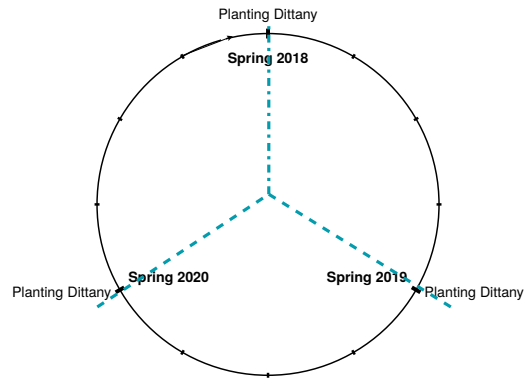

(c) A recurrent dataset on a circle ○ timeline.

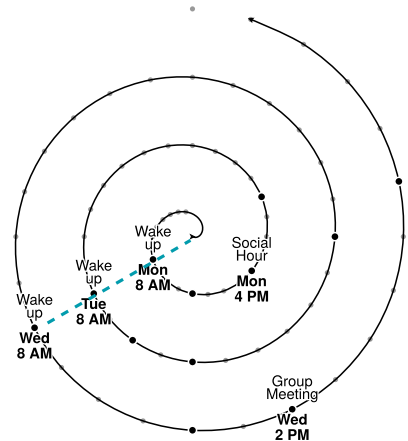

(d) A mixed dataset on a spiral ◉ timeline.

Figure 2. Examples of the stimuli used in the experiment. All the datasets have been simplified to aid readability within the paper.

relative angle from center aligns periodic activities along an axis. As shown in Figure 2dfor a mixed dataset, a recurring pattern 'Wake up' aligns along an axis. The final experimental stimuli are available at https://osf.io/qr5yu/

**Experiment**

Participants. We recruited 192 participants. The experiment was expected to last 5 minutes, and each participant received $2 base pay plus a bonus of $0.2 for each correct question after the sixth correct one. Our participants self-reported their demographics.. By gender 53% were male, 31% were female, while the rest chose not to answer. By education 32.8% had a Bachelor's degree, 17.7% had a high school degree, 16.6% attended some college but do not have a degree, and the rest had either a Master's degree, an Associate Degree, a Doctorate, had no degree, chose not to answer. By employment 62.5% were employed for wages, 17.2% were self-employed, 5.2% were out of work, and the rest were either students, homemakers, retired, unable to work, or chose not to answer.

Experiment Tasks. In order to determine the user tasks most relevant for timelines we interviewed a professor working in the history field and a PhD student with personal health data from Northeastern University, Boston. Both of them were selected because their research dealt with timelines and sequences of temporal information. The semi-structured interviews lasted about 1 hour. We asked them to discuss the information that is important to show on a timeline and what kind of tasks would be relevant for the use of timelines in their own field. We based our task analysis on the information we obtained from the interviews using Brehmer and Munzner's task taxonomy [9]. If we say that the target of a task is the name of an event, and that the year is the position, we can formulate specific questions that correspond to tasks. A locate task, for example, can be formulated on a timeline containing daily events as "At what time was lunch on Monday?" In this way, the question gives the target (lunch) and asks for the location (at what time?).

- **When** did <event> happen?
  *Example: When did the earthquake happen?*
  In this case, we are asking for the location of a target. The location is unknown but the target is known. Users would need to explore the timeline and report the year as an answer.
  Tasks: **Discover** → **Locate** → **Identify.**

- **What** happened at <time>?
  *Example: What happened in 1999? What happens at the start of the timeline?*
  Location is given, target is not. Users will look at the location and report the name of the event.
  Tasks: **Discover** → **Browse** → **Identify.**

- **Find** <event> that happened at <time>?
  *Example: The earthquake happened in 1898. Click on it.*
  Participants will answer this type of question by clicking on a specific point on the timeline. This question gives to the users both the target and the location.
  Tasks: **Discover** → **Lookup** → **Identify.**

- **Compare** time of <event1> relative <event2> or <event3>?
  *Example: Did Cleopatra live closer to the launch of the first iPhone or the construction of the Pyramids?*
  Users need to find 3 data points on the timelines and then compare the distance between pairs of data points.
  Tasks: **Discover** → **Explore** → **Compare.**

Experiment Design. Upon approval from the Institutional Review Board of Northeastern University, we recruited participants from MTurk. In order to be able to use the platform, we developed a web application to run the test directly on the participants' home computers. After running an initial pilot with 16 participants, we used a power analysis (see below) which indicated that we needed to recruit 192 participants. They were randomly but equally divided into 4 main groups, and each group of participants was assigned questions belonging to a particular task to be performed on a timeline (i.e., group 1 only performed the "when" task, group 2 only performed the "what" task, etc.). In order to minimize learning and ordering effects, we used Latin squares to determine the sequence of datasets and timeline shapes that the participants were shown.

Experiment Apparatus. MTurk was used to recruit and pay the participant, but the experiment was actually hosted on our own domain. It was served as a web app in which the participants could interact with both keyboard and mouse to go through the tutorial and answer the questions. All the code was written in JavaScript, ran on the participant's own device, and it rendered the stimuli and questions at run time in svg in the participant's browser using their screen size to determine the size of the visualization.

Experiment Procedure. After accepting the HIT, participants were asked to agree to our informed consent. They were then asked to complete a short tutorial that introduced them to the different timeline shapes used in the test and how to interact with the web application. The test required each participant to answer 12 questions. Groups 1, 2, and 4 answered multiple choice questions while group 3 was asked to click on timelines to find the answers. At the end, participants were prompted to answer an optional survey (described in the next section). After completing the study participants were given a unique code that they had to copy and paste into a prompt on MTurk so that they could get paid based on their performance.

Survey. The participants were given a short optional survey upon completion of the 12 questions. The survey contained questions about the demography of the participants, their feedback and whether they used any strategy while answering the questions. We also asked the 4 groups of participants which timeline shape was the most readable for each of the 3 different datasets. Readability here denotes which shape was the easiest to follow and read given a dataset. This was included to get an idea of which shapes worked best with what kind of dataset. The reason behind asking the "strategy" question was to make sure that the participants used the timeline shapes to navigate themselves in finding the correct answer and not some other strategy that depended less on the shape.

**Data Analysis**

To analyse the results of our experiment we used null hypothesis significance testing as well as interval estimation of effect size [15]. The analysis plan was pre-registered be-

fore we conducted the experiment and is available at **https://osf.io/qr5yu/**

**Required Sample Size Based on Power Analysis.** In order to effectively test for the effects of time and accuracy across the different timeline shapes, we conducted a power analysis to estimate the number of participants needed to identify a statistically significant difference with a Type I error rate $\alpha$ of 0.05. We used the G*Power [16] for power analysis. G*Power does not support Friedman's non-parametric repeated measures test to compare two or more samples. As an alternative, we estimated the sample size by comparing two samples using the one-tailed Wilcoxon signed rank test. Since we estimated the sample size pairwise we used conservative input parameters to handle the multiple comparisons and to estimate the upper bound on the number of participants required for the study. The parameters used in the study were Type I error rate $\alpha$ of 0.83% (Bonferroni corrected) at a power level of $1 - \beta$ of 95%. The effect size was individually calculated using the mean completion time from the pilot study. The number of participants was estimated to be 48 for each task when rounded up to the nearest multiple of 12 to ensure that each group in the Latin square had the same number of participants. Thus a total of $48 \times 4 = 192$ participants were recruited.

**Outlier Removal.** Before proceeding with statistical tests comparing the time and accuracy distributions of different timeline shapes we identified a sizable set of data points that had particularly large measurements of time. Such outliers are quite common in time-based measurements from user studies and, if included without any special consideration, they can severely affect the quality of the statistical analysis. Outliers in the time data are usually not attributed to a group of participants that found a question excruciatingly difficult but most likely they were distracted during that question. In the proceeding analysis we considered a time data point for a given task as an outlier if its value is two or more standard deviations greater than the mean time for its given task. Across all 4 tasks we classified a total of 66 time data points as outliers and they originated from 47 participants. Those outlier data points were not part of any of our statistical analysis.

**Time Analysis.** In order to ensure that our statistical tests would not violate any of their necessary assumptions we examined the time distributions per dataset for each (task, timeline shape) pair. In particular, we tested for the normality of each time distribution using Q-Q plots and the Shapiro-Wilk test [31] and interpreted the $p$-values holistically rather than dichotomously. We determined that our time data was not normally distributed. We also performed Box-Cox transformations [29] for each time distribution and not all of them were transformable to a normal distribution using the same exponent, thus we considered *non-parametric statistical tests* that don't assume a normal distribution of the data.

In order to examine if there was a difference in the time distributions between the 4 timeline shapes for a given dataset we ran a Friedman test [17] comparing the 4 distributions with the null hypothesis that all timeline shapes have the same time distribution. A Friedman test is an appropriate choice as it is the *non-parametric equivalent* of a one-way ANOVA test with repeated measures. Notice that each participant in a given task saw — i.e., repeated — each timeline shape 3 times. If the null hypothesis of the Friedman test was rejected that implies that there was at least one pair of timeline shapes whose time distributions were different at the $\alpha = 5\%$ level. To identify which pairs of timeline shapes were different we ran a Nemenyi test [24] for each dataset as our post-hoc analysis. The Nemenyi test performs all pairwise comparisons between our different timeline shapes and the $p$-values it reports for each pair are adjusted for multiple hypotheses testing. We then interpreted these $p$-values more holistically in the context of our interval estimates (see below).

To examine if there is difference in time between the 4 timeline shapes for a given (task, dataset) pair we ran a Friedman test [17] for each pair with a Nemenyi test for the post-hoc analysis just like we did for our per-dataset analysis as we explained previously. Notice, because the Friedman test requires that each participant to have a data point in each distribution (i.e., timeline shape) we sometimes run into the problem of missing data due to the removal of outlier time data from our analysis. E.g., if a participant's time performance for the spiral timeline was classified as an outlier we cannot run the Friedman test using that participant's data even though we have their time performance on the other timeline shapes. As a result, workers whose data included an outlier were pruned from the Friedman test. It's important to emphasize however that the number of workers pruned for each group that the Friedman test was applied to was relatively small at around 0–15% of the total number of participants in the group being tested.

**Accuracy Analysis.** Because each participant sees only one question per (dataset, timeline shape) pair their accuracy for a given pair is either 100% (correct) or 0% (incorrect). As a result, our accuracy analysis resorts to *non-parametric statistical tests* due to the binary nature of our data distribution. In particular, we use a Chi-Square test for independence [23] for each (task, dataset) pair with the null hypothesis that there is no association between timeline shape and the response distribution. In other words, the null hypothesis is that user response (i.e., accuracy) is not dependent on the timeline shape. Just like our time analysis interpreted the $p$-values holistically rather than dichotomously.

**Interval Estimation of Effect Size.** Although null hypothesis significance tests are ubiquitously used to test for the validity of alternative hypotheses, they can also be problematic due to their dichotomous nature [4] and their imposed arbitrary $p$-value $< .05$ threshold that's used for denoting scientific findings of statistical significance. In addition to our statistical tests with null hypothesis significance testing, we also report effect sizes with interval estimates [15] in order to uncover more nuanced trends in our data that cannot be represented by just a $p$-value. To do so we provide the 95% confidence intervals (CIs) via bootstrapping[1] for each (task, dataset) pair to indicate the range of plausible values of the mean completion time and mean proportion of correct responses [12]. We also examine the 95% CI for the mean per-worker log change in completion time [34]) using the the most well-known timeline shape (linear horizontal) as the comparison baseline. Similar examples of using estimation for this type of analysis are presented in [3, 8]. Note one exception to our pre-registration: we had planned to use the ratio of completion time instead of the log change in completion time but discovered that the former is asymmetric [34].

---
[1] Bootstrapping involves randomly drawing observations from the experimental data with replacement in order to assess many alternate datasets, and thus use the variability across these datasets as a proxy for sampling error [15].

|        | When |      |      |      | What |      |      |      | Find |      |      |      | Compare |      |      |      |
|--------|------|------|------|------|------|------|------|------|------|------|------|------|---------|------|------|------|
|        | C    | LH   | LV   | S    | C    | LH   | LV   | S    | C    | LH   | LV   | S    | C       | LH   | LV   | S    |
| **Mixed** |   |      |      |      |      |      |      |      |      |      |      |      |         |      |      |      |
| C      |      | 0.59 | 0.90 | 0.90 |      | 0.39 | 0.00 | 0.90 |      | 0.06 | 0.01 | 0.83 |         | 0.90 | 0.20 | 0.90 |
| LH     | 0.59 |      | 0.64 | 0.90 | 0.39 |      | 0.05 | 0.63 | 0.06 |      | 0.90 | 0.32 | 0.90    |      | 0.47 | 0.90 |
| LV     | 0.90 | 0.64 |      | 0.90 | 0.00 | 0.05 |      | 0.00 | 0.01 | 0.90 |      | 0.12 | 0.20    | 0.47 |      | 0.24 |
| S      | 0.90 | 0.90 | 0.90 |      | 0.90 | 0.63 | 0.00 |      | 0.83 | 0.32 | 0.12 |      | 0.90    | 0.90 | 0.24 |      |
| **Non-Recurrent** | | | | | | | | | | | | | | | | |
| C      |      | 0.53 | 0.71 | 0.01 |      | 0.90 | 0.03 | 0.66 |      | 0.36 | 0.36 | 0.90 |         | 0.90 | 0.90 | 0.06 |
| LH     | 0.53 |      | 0.90 | 0.35 | 0.90 |      | 0.01 | 0.79 | 0.36 |      | 0.90 | 0.12 | 0.90    |      | 0.90 | 0.01 |
| LV     | 0.71 | 0.90 |      | 0.20 | 0.03 | 0.01 |      | 0.00 | 0.36 | 0.90 |      | 0.12 | 0.90    | 0.90 |      | 0.02 |
| S      | 0.01 | 0.35 | 0.20 |      | 0.66 | 0.79 | 0.00 |      | 0.90 | 0.12 | 0.12 |      | 0.06    | 0.01 | 0.02 |      |
| **Recurrent** | | | | | | | | | | | | | | | | |
| C      |      | 0.83 | 0.79 | 0.00 |      | 0.47 | 0.71 | 0.56 |      | 0.90 | 0.89 | 0.33 |         | 0.90 | 0.90 | 0.03 |
| LH     | 0.83 |      | 0.90 | 0.01 | 0.47 |      | 0.90 | 0.03 | 0.90 |      | 0.74 | 0.21 | 0.90    |      | 0.90 | 0.05 |
| LV     | 0.79 | 0.90 |      | 0.01 | 0.71 | 0.90 |      | 0.10 | 0.89 | 0.74 |      | 0.74 | 0.90    | 0.90 |      | 0.02 |
| S      | 0.00 | 0.01 | 0.01 |      | 0.56 | 0.03 | 0.10 |      | 0.33 | 0.21 | 0.74 |      | 0.03    | 0.05 | 0.02 |      |

Figure 3. Exact *p*-values for the time analysis from the Nemenyi post-hoc test for the 4 tasks (columns) and 3 data types (rows). Low *p*-values indicating differences are highlighted in orange. Note that the matrix is symmetric.

**Qualitative Data Analysis.** In order to get a sense of which timeline shapes did well in terms of readability across all the datasets in each the 4 tasks (When, Where, Find, and Compare), we simply calculated the total number of responses for readability for each category (horizontal line, vertical line, circle, and spiral) and found their percentages using the total number of responses per task. The results are discussed later in the paper. To analyze the comments left by the participants we used open coding [21]. In particular, we applied conventional content analysis where the categories of code are derived directly from the text data [19].

## RESULTS

**Quantitative Results.** The results of our study are summarized in Figure 4, which shows the mean completion time, mean log change in completion time, and mean proportion correct for each combination of task, dataset, and timeline shape. Moreover, the results of pairwise Nemenyi tests, from our post-hoc analysis of the Friedman tests, are shown in Figure 3. We identified the following meaningful differences in time between pairs of timelines shapes for a given (task, dataset) pair. We use the timeline glyphs: ○ , − , | , ◉ to represent circle, linear horizontal, linear vertical, and spiral timeline shapes, respectively. The < symbol between timeline glyphs indicates that the mean time of the shape on the left is likely meaningfully smaller than the mean time of the shape on the right. The = symbol indicates likely similar mean times.

**When Task.**
- Mixed data: ○ = | = ◉ ($p = .90$ for all) and − = ◉ ($p = .90$).
- Non-recurrent data: ○ < ◉ ($p = .01$) and − = | ($p = .90$).
- Recurrent data: ○ < ◉ ($p < .001$), − < ◉ ($p = .01$), | < ◉ ($p = .01$), and − = | ($p = .90$).

**What Task.**
- Mixed data: | < ○ ($p = .01$), | < − ($p = .05$), | < ◉ ($p < .001$), and ○ = ◉ ($p = .90$).
- Non-recurrent data: | < ○ ($p = .03$), | < − ($p = .01$), | < ◉ ($p < .001$), and − = ○ ($p = .90$).
- Recurrent data: − < ◉ ($p = .03$) and − = | ($p = .90$).

**Find Task.**
- Mixed data: | < ○ ($p = .01$) and − = | ($p = .90$).
- Non-recurrent data: ○ = ◉ ($p = .90$) and − = | ($p = .90$).
- Recurrent data: ○ = − ($p = .90$) and ○ = | ($p = .89$).

**Compare Task.**
- Mixed data: ○ = − = ◉ ($p = .90$ for all).
- Non-recurrent data: − < ◉ ($p = .01$), | < ◉ ($p = .02$), and − = | = ○ ($p = .90$ for all).
- Recurrent data: ○ < ◉ ($p = .03$), − < ◉ ($p = .05$), | < ◉ ($p = .02$), and − = | = ○ ($p = .90$ for all).

Through our accuracy analysis we failed to find any meaningful differences between any pair of timeline shapes for any given (task, dataset) pair. Details at **https://osf.io/qr5yu/**

**Qualitative Results.** As part of the survey, participants were asked to select the timeline shape they perceived as the easiest to read for each dataset. Figure 5 shows the results. The horizontal line − was largely chosen as the easiest to read, especially for the non-recurrent dataset. Responses where more mixed for the recurrent dataset where, depending on the task, circle ○ and spiral ◉ were selected more.

We also received some interesting comments that pertained directly to the shape of the timelines. This gave us a notion of how the participants felt about certain shapes to an extent that they cared enough to express it in an optional feedback section. In this case too the majority of the comments were in favor of the horizontal and vertical lines. The comments were of the similar trend: *"...the vertical line is easy to read"*, *"The horizontal and vertical graphs were easiest"*, etc. A few remarked in favor of the circular shape: *"...but circles were easiest for my eyes"*. The spiral shape also had a few admirers under its belt with comments like: *"The planting dataset with the spiral is a good example of a timeline that is functional and visually appealing"*. However, on the flip side, some of these participants revealed concerns about the spiral shape, stating it to be difficult and confusing. Their comments echoed similarly: *"I found the spiral hard to work with"*, *"...the spirals were confusing"*, *"The spiral one was really hard"*.

## DISCUSSION

In this paper, we present a novel study comparing timeline shapes. Our participants completed a controlled visualization experiment and we measured their performance in terms of time and accuracy for completing analytical tasks. Although we hypothesized that timeline shapes would not affect a person's performance, our results suggest the opposite, and we found timeline shape does indeed affect how people read a timeline visualization. In this section, we discuss the relevance of our experimental design and quantitative (Accuracy and Time) and qualitative user performance measures. After establishing the significance of our experimental design, we analyze the results and discuss the implications of findings on the design principles for timelines.

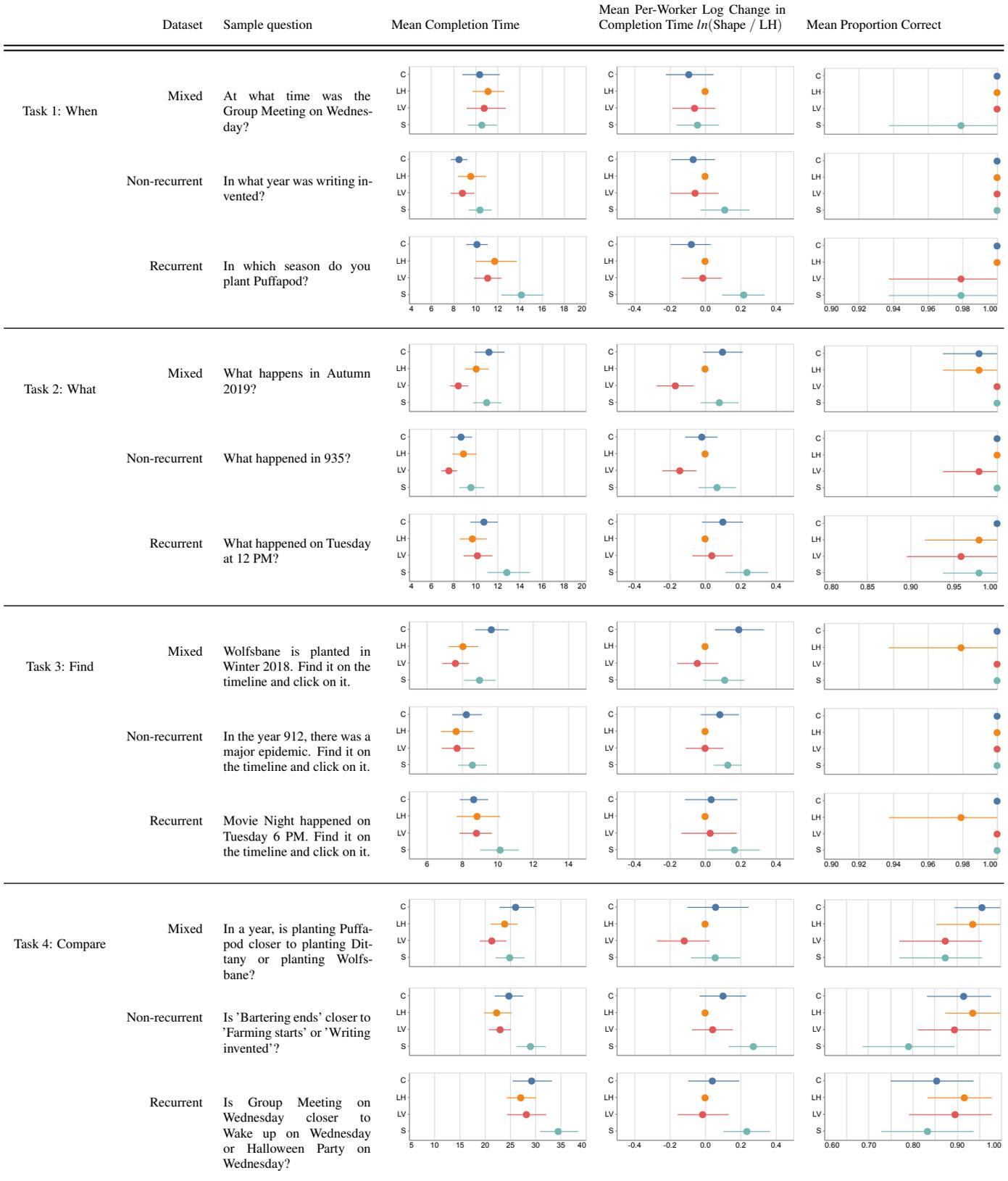

Figure 4. Quantitative experiment results for the 4 experimental tasks, each with 3 datasets and 4 timeline shapes — circle (C, ●), linear horizontal (LH, ●), linear vertical (LV, ●), and spiral (S, ●). For reference, we show a sample question for each task-dataset pair. Error bars show the bootstrapped 95% confidence interval (CI) of the mean [12, 15]. Left: Mean completion time. Middle: Mean per-worker log change in completion time using each worker's linear horizontal (LH) time as the baseline. Note that natural log ratios are the only symmetric, additive, and normed indicators of relative change [34]. Right: Mean proportion of correct responses. In the left and right columns the horizontal axis scale is relative for each task.

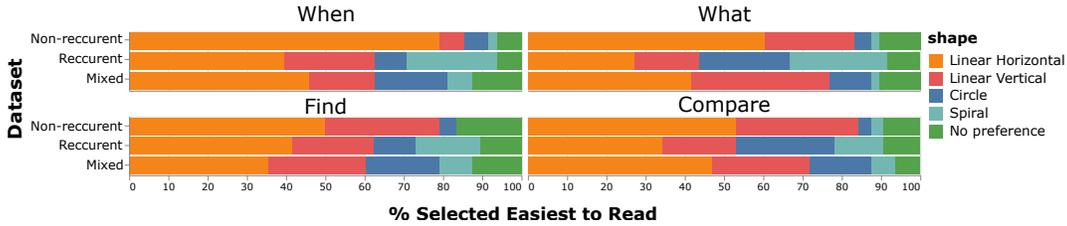

Figure 5. Comparison between the responses of the users when asked which shape was the easiest to read for each one of the tasks.

**Reflection on the Study Design.** The design space of timeline shape and the ways to interact with them has grown rapidly over the last few decades [6, 30]. However, evaluation of this design space is still in the early stages (see Related Work). In this experiment, we evaluate the primary design factor of timelines, their shape. To evaluate the shape, we try to minimize all other factors to get the best estimation of the effect of timeline shape. Therefore, we chose a study design where the timelines were non-interactive with consistent design aspects such as the display aspect ratio, the font-size of tick labels, etc. as discussed in the Stimuli section. Our evaluation of minimalist timelines raises a question of the applicability of these results to timelines designed with complicated interactions or higher data density. We argue that a minimalist design also minimizes the chances of eliciting a difference in user's performance in the study. In other words, in a simple experiment, the error rate and total time to complete the task should be similar across the timeline designs. Therefore, if we notice any differences with a minimalist design, then the chances of noticing the differences should increase with the complexity of design and scale of data, unless the design is prepared to handle the scale of data specifically. The same argument can be made for the simplicity of the chosen tasks: many taxonomies of low-level tasks, e.g., the one we use [9], argue that sensemaking is built upon combinations of such low-level tasks.

Visualization tasks may require a varying level of cognitive effort. E.g., a task where a user has to find the time of an event (When task) is cognitively easier than a task to compare times for two events (Compare task). The dual-process of decision making [26] suggests that hard cognitive tasks may be slower to perform as they require more contemplative decision making. To reduce variation in task performance introduced by task complexity we use clear compartmentalization of tasks. Our study measures the difference in timeline readability with different timeline shapes. We argue that the total time and accuracy of a user with a timeline gives evidence about its readability. Quantitative measures do not provide a holistic view of the user perception of the readability of the timeline design and user perception and sentiment are important to understand. Consequently, we survey the perceived readability of a visualization design at the end of the experiment. We use both the quantitative and qualitative measures to predict the overall efficacy of timeline shapes for representing temporal event data.

**Effect of Shape on Timeline Readability.** In general, we found that participants are faster at reading information from linear − | timelines vs. circles ○ and spirals ◉ . In 6 out of 12 comparison conditions, shown in Figure 4, either one or both of the linear timelines were meaningfully faster than the circle or spiral timelines. We also found linear shapes were as perceived more readable. A valid explanation for these results can be our participants' familiarity with linear timeline designs. In their survey of timeline visualizations Brehmer et al. [6] found 73% of timelines surveyed used a linear representation. Familiarity with the timeline design may also imply that participants had prior training to read the linear timelines or were faster to adapt to the linear design.

We found substantial evidence that participants were slower at reading the spiral ◉ timeline and also perceived it as less readable. In 7 out of 12 conditions, shown in Figure 4, spiral timelines were the slowest or on par with the slowest. These results are surprising, as prior research [11] presented evidence that spiral timelines work particularly well for representing serial-periodic data. We attribute the unfavorable results for spiral timelines to its lack of familiarity. Lack of familiarity with the timeline shape can make even a simple visualization task cognitively hard and lead to slow task completion times.

We found strong evidence that timeline shape affects task completion time. It is important to note that the time differences between timeline shapes were small in absolute values mean time differences ranging from 1–7 seconds (see Figure 4). However, this equated to a mean log change in completion time ranging ± 20 L% from the mean time for a linear horizontal timeline. We believe the main reason for this meaningful but relatively small effect size is the overall experiment design. We designed straightforward experiment stimuli, scaled the datasets to make the stimuli less overwhelming, and used common visualization tasks. As a result, we argue our experiment was not very difficult to complete for the participants. Some participants also acknowledged this in their feedback: *"All of the timelines were quite easy for follow and understand"*, *"The timelines were all very easy to read no strategy required"*. As a result we observe relatively small effects for time differences. However, posit that these result trends should hold generally and the effect size will increase as the task becomes harder.

Our evidence suggests timeline shape does not meaningfully affect user accuracy (see Figure 4). No general patterns seem to emerge from the results for the When, What, and Find tasks (see Figure 4). However, we can not conclude that the mean accuracy is completely similar either. As with the time effect size, our straightforward experiment design may be the reason for these accuracy results. In the compare task (see Figure 4), we notice a tendency for people to make more errors with spiral ◉ timelines. However, the results are not different enough to make any definitive claims.

**Effect of Task on Timeline Readability.** We never intended to compare the results across the different tasks. However, the difference in completion times and accuracy among the tasks may have real-world design implications. We found good evidence that participants were slower and less accurate with all the timeline shapes in the Compare Task of the study (see Figure 4). We argue this happened because the Compare task is cognitively harder than the other tasks. Furthermore, spirals  may be cognitively harder to process, and using them for comparison tasks may compound task difficulty. As a result, we suggest designers use a simple timeline design like the linear design  when the task is complicated, unless the task specifically requires the use of spiral timelines.

**Effect of Dataset on Timeline Readability.** In this study, we compare the 4 timeline shapes across 3 datasets. We do this to measure the association between timeline shape and dataset type. However, in our study we did not find any association between timeline shape and datasets. We thought the linear timelines  would be more readable with linear data, while circle  and spiral  timelines will be easier to read with recurrent and mixed datasets. We did not find enough evidence to support this claim. Based on these results, designers have the flexibility of representing linear data on spiral timelines to make the visual design enjoyable — with the caveat of the expressiveness criteria discussed in Design Recommendations below. Vice versa, a complex mixed dataset can be represented with a linear timeline to aid faster readability.

**Usability of Non-Linear Timelines: Circle and Spiral.** We found that linear timelines  are faster to read. However, usefulness and timeline readability are two different measures. There might be situations where non-linear timelines may be more informative, metaphorically similar to the dataset, or simply pleasing to eyes. In that scenario, the designers may choose aesthetic appeal over optimal speed. An important readability measure that goes in favor of non-linear timelines is the lack of difference in accuracy for cognitively easy tasks. If the consideration for readability is just accuracy, then we suggest liberal use of the timeline shapes — again with the caveat of the expressiveness criteria discussed in Design Recommendations below.

## DESIGN RECOMMENDATIONS
Here we present a list of design recommendations derived from our results. We chose these recommendations based on their broad applicability. Our design recommendations are targeted at suggesting users the right design if their primary concern is to increase the readability of temporal event sequence data. However, based on domain and context, these recommendations may not directly be applicable. In such a case, we request designers to practice caution while using these recommendations.

**Context, Usability, and Expressiveness.** First of all, the importance of differences in timing are dependent on the context. E.g., in emergency medicine a difference of a few seconds may be much more relevant than for consuming a timeline published in a magazine. It is also important to consider the expressiveness of the data — how well the timeline shape can represent the underlying data. For instance, representing linear data on a circular shape may be misleading to the readers by inducing them into thinking there are recurrences in the data when there are none. We therefore suggest designers evaluate our recommendations based on their domain goals and specific use case.

**Timeline Readability.** Based on our results we formulate the following recommendations:

1. Use linear vertical timelines  for situations which require fast data lookup.
2. Avoid spiral timelines  when the task requires fast lookup.
3. If you use a spiral timeline  also include a tutorial or visual cues to assist the user in learning and understanding.

**Task Performance.** Tasks which require long-term memory dependence like the Compare task are seem slower and less accurate on all the timeline designs. This effect is amplified with spiral timelines . We recommend using linear timelines  for difficult tasks which require complicated decision making.

**Dataset Flexibility.** Within this study, we did not find that dataset choice affects the readability of the timeline. Therefore, we recommend designers to be flexible with their choice of timeline shape to maximize readability or improve engagement. However, if the dataset is complex, even for mixed data, we recommend using a linear timeline .

## LIMITATIONS AND FUTURE WORK
In this study we measure the effect of shape on timeline readability. Besides shape, several other factors influence how people read timelines. Factors like data density, the use of interaction, visual embellishments like pictures on the timeline, and the difference in path length between a shape and another may also affect timeline readability. Future work can study the additional design characteristics proposed by Brehmer et al. [6] like scale (log or relative) as well as additional layouts (e.g., multiple stacked lines). The timeline design space is vast and we believe there is extensive room for future work to help us understand timelines better.

## CONCLUSION
Timeline visualizations are pervasive in our everyday life. However, we know little about ways to design timelines effectively. We present a novel study to measure the effect of 4 common timeline shapes on timeline readability: a horizontal line , a vertical line , a circle , and a spiral . From our results, we learned that timeline shape affects the readability of timelines. More specifically, we found good evidence that the linear shapes support reading the timelines more quickly than the non-linear shapes. We also found evidence that non-linear spiral shape is not only perceived slowest by users but also leads to slower lookup of events with timelines. Future studies and real-word timeline designers should carefully analyze their data representations in light of our findings.

## ACKNOWLEDGEMENTS
We thank the National Science Foundation for support, in part, under CRII award no. 1755901 and our reviewers & colleagues for their advice.